# A Mathematical Model for Simulating Meteor Showers

M. CARDINOT[1] and A. NAMEN[2]

**ABSTRACT.** This paper presents a mathematical model to simulate the trajectory of a meteor as seen by a single observer located anywhere on Earth. Our strategy is to define a new coordinate system, called Radiant Coordinate System, which is centered on the observer and has its z-axis aligned with the radiant. This new coordinate system allows us to describe the meteors' path by applying a reduced number of equations in a simple solution. We also present a computational implementation of this model, which is developed as a new plug-in of *Stellarium*, a free and open-source planetarium software. Moreover, we show that our model can be used to simulate both meteor showers and sporadic meteors. In particular, meteor showers are simulated using data provided by real catalogs.

## 1 INTRODUCTION

When looking at the sky for a few minutes, it is often possible to observe some streaks of light which are popularly known as "shooting stars" or "falling stars". Contrary to what these names suggest, such streak of light is a meteor and not a star. Occasionally, we can observe a *meteor shower*, which is a phenomenon where a greater number of meteors seem to originate from the same point in the sky at a specific period. This point is known as *radiant* and is usually labeled according to the nearest bright star or constellation [8]. Meteor showers attract people's attention mainly because of its periodicity and for being one of the few celestial events which can be seen with the naked eye, i.e., without the need of optical instruments such as binoculars, telescopes and spotting scopes.

This paper proposes a mathematical model that allows a better understanding of how a sporadic meteor or a meteor shower may be seen by a single observer on Earth. Although this problem has a simple formulation, it refers to a three-dimensional scenario (i.e., the displacement of a meteor can occur in all directions in the sky), which might be difficult to be modeled. Furthermore, this work aims to implement the proposed model in a free and open-source planetarium software called *Stellarium* to enable users to find and track radiants in a virtual three-dimensional sky.

*Corresponding author: Marcos Cardinot – E-mail: marcos@cardinot.net – https://orcid.org/0000-0001-6566-0139

[1] Department of Information Technology, National University of Ireland Galway, University Road, H91TK33, Galway, Ireland. E-mail: marcos@cardinot.net

[2] Instituto Politécnico, Universidade do Estado do Rio de Janeiro, Rua Bonfim 25, Vila Amélia, 28625-570, Nova Friburgo, RJ, Brazil. Universidade Veiga de Almeida, Rua Ibituruna 108, 20271-901, Maracanã, RJ, Brazil. E-mail: aanamen@uol.com.br

There are certainly many ways to tackle the problem of formulating a mathematical model to understand the meteor dynamics in the Earth's atmosphere. However, as we are interested in implementing this solution in a computer software, it must be as simple and functional as possible. Taking advantage of the Earth's rotation and considering that the position of a radiant in the celestial sphere is known, our strategy is to define a new coordinate system, called Radiant Coordinate System (RCS), which is centered on the observer and has its $z$-axis aligned with the radiant.

In this work, we show that the Radiant Coordinate System allows us to develop a computational model performing a small number of calculations, which in turn, is able to handle very high zenithal hourly rates (ZHR) for any meteor shower at a very low computational cost. Moreover, the model considers that meteor showers can be active at different locations on Earth and at the same time.

The paper outline is as follows: Section 2 presents an overview of the background information necessary to understand some of the characteristics and behavior of meteors, which in turn, provide the basis for the development of the mathematical model presented in sections 4 and 5. In Section 3, we describe the RCS, which simplifies the model and the development process. Section 4 presents the equations used to describe the meteor path. Section 5 presents the modeling of the Zenithal Hourly Rate. In Section 6, we discuss the results from the computational implementation of the model in a free and open-source software. Finally, in Section 7, we present our conclusions and outline future work.

## 2 BACKGROUND

The modern relationship between comets and meteors and their composition has been studied and refined for a long time [11], and it was first referred as a "dirty snowball" by Whipple in his seminal work [19]. Considering that a comet consists of an icy nucleus made up of dust grains in various sizes. As a comet approaches the Sun, it gradually sublimates releasing the dust grains which can form a tail of tiny grains and gases, and a meteoroid stream. The meteoroid stream will move on a similar orbit to the comet. When the Earth passes through such a stream, some of these meteoroids penetrate the Earth's atmosphere with a velocity that ranges from 11 to 72 $km/s$ [17, 8, 12].

Meteors are produced when a meteoroid, i.e., a piece of stony or metallic body in outer space, interacts with the Earth's atmosphere. In this process, air molecules collide with the meteoroid, in which the transfer of momentum and energy heats up the meteoroid. As it heats up, material ablates away and further collides with air molecules, which creates a column of ions, electrons, and photons, known as a meteor gas trail.

## 2.1 Meteor's altitude

In general, meteors will be visible in the region called thermosphere, ranging from 80 to 120 *km* in altitude [8]. However, it is noteworthy that the initial and final height also depends on the meteor speed, i.e., the faster the impact on the Earth's atmosphere, the higher the altitude; on the other hand, slower meteors will appear in a lower altitude. Moreover, other aspects and physical properties may also affect the meteor altitude, such as the meteoroid mass, its entry angle and the atmospheric density.

## 2.2 Meteor's distance from an observer

According to Richardson [16], considering that meteors occur in an altitude ranging from 80 to 120 *km* (see Subsection 2.1) and that Earth's equatorial radius is approximately 6378 *km*, ignoring topographic irregularities, it is possible to construct a cross-sectional view of the area in which meteors occur for a single observer on the Earth's surface. Fig. 1 shows that the meteor distance from the observer can be obtained by applying the law of cosines to the triangle formed by the meteor, the observer and the center of the Earth, obtaining:

$$(r+h)^2 = r^2 + d^2 - (2rd\cos(\pi - \zeta))$$

which can be solved to find the meteor distance from observer ($d$),

$$d(h, \zeta) = \sqrt{r^2\cos^2\zeta + 2rh + h^2} - r\cos\zeta \qquad (2.1)$$

where $r = 6378$ *km* corresponds to the Earth's equatorial radius, $h = [80, 120]$ *km* is the meteor altitude and $\zeta$ is the meteor zenith angle. Note that when $\zeta$ is less than 65°, it might be more convenient to use the following equation:

$$d(h, \zeta) = \frac{h}{\cos\zeta} \qquad (2.2)$$

which is a first order approximation of 2.1. Fig. 2 shows meteor distances ($d$) for a meteor with an altitude of 120 *km* as a function of meteor zenith angle ($\zeta$) for Equations 2.1 and 2.2. In fact, both equations agree well for angles below 65°.

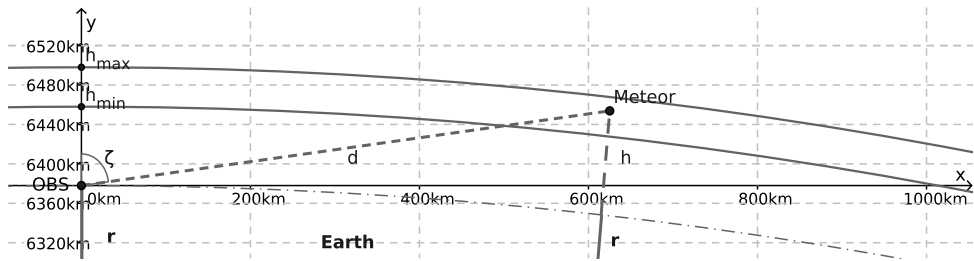

Figure 1: A cross-sectional view of the area in which meteors occur. In this view, $\zeta$ corresponds to the meteor zenith angle; $d$ is the meteor distance from the observer; $h$ is the meteor altitude and $r$ corresponds to the Earth's equatorial radius. Extracted from [3].

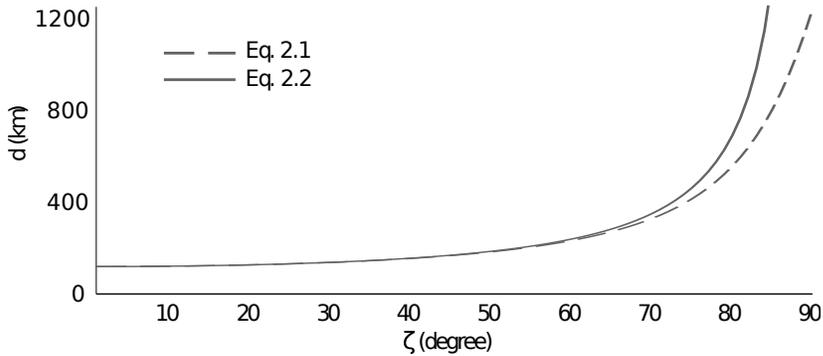

Figure 2: Comparison between Equations 2.1 and 2.2 for a meteor at an altitude of 120 $km$ ($h = 120\ km$).

Note that as $h = [80, 120]$, from Equations 2.1 and 2.2, it is possible to find a maximum-minimum distance between the meteor and the observer by only knowing the meteor zenith angle $\zeta$.

## 2.3 Zenithal Hourly Rate

In a meteor shower, the Zenithal Hourly Rate (ZHR) refers to the number of meteors which would be seen per hour by an observer on the peak day, under perfect sky conditions, i.e., to a limiting magnitude of 6.5, and with the radiant in the zenith [1].

Hence, as the radiant is rarely at the zenith and the sky conditions are usually non-perfect, the effective ZHR is lower than the rate that we find in the catalogs. In reality, the closer the radiant is to the horizon, and as we move away from the peak day, lower is the effective ZHR. Note that for an observer, knowing the Zenithal Hourly Rate of a meteor shower is as important as knowing the coordinates of its radiant.

## 3 RADIANT COORDINATE SYSTEM

The choice of the coordinate system is undoubtedly a crucial task in the description of a physical problem. A convenient and well thought-out choice may facilitate the understanding of the problem as it can reduce the number of equations involved and, consequently, simplify the solution.

As previously discussed, we know that the position of the radiant in the sky is one of the most important information about a meteor shower, because it allows the observer to know from which region of the sky the meteors will originate. Obviously, viewing a meteor shower also depends on the date and location of the observer on the Earth's surface. Thus, as we take the observer into account, the use of the Horizontal Coordinate System (HCS) seems to be convenient or necessary at some point. Also, although the majority of the meteor shower calendars inform the position of the radiant in the Equatorial Coordinate System (ECS), the conversion between the ECS and the HCS is widely known [5, 14, 13]. Therefore, as the starting point for this model, we assume that the position of the radiant in the HCS is known.

Thus, as illustrated in Fig. 3, the position of any radiant in the HCS is given by the altitude $\lambda$ and the azimuth $\phi$ angles. Here we also introduce the components $x$, $y$ and $z$ in the HCS such that: it is centered on the observer, $z$-axis is aligned with zenith, $x$-axis is aligned with the south pole, and $xy$ plane is perpendicular to $z$. Notice that the $y$-axis will consequently be aligned with East. Thus, the azimuthal angle will be measured from the South increasing towards the East, which corresponds to the convention adopted in the software *Stellarium*.

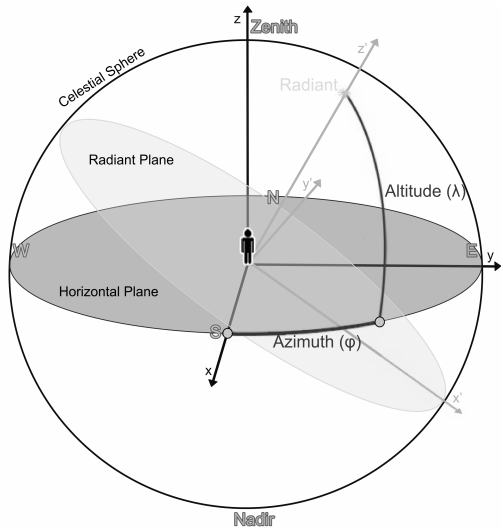

Figure 3: The Radiant Coordinate System, which is obtained from two successive rotations of the Horizontal Coordinate System.

We aim to develop a model which describes the displacement of a meteor in the celestial sphere with respect to an observer located anywhere on Earth's surface. In order to facilitate this modeling, we propose the creation of a new coordinate system i.e., the Radiant Coordinate System (RCS), which has the following characteristics: it is centered on the observer, $z'$-axis aligned with the radiant, and $x'y'$ plane is perpendicular to $z'$-axis. In fact, the RCS resembles the HCS, where the main difference is that the RCS is configured to have the $z'$-axis aligned with the radiant rather than the zenith.

It is important to note that due to the Earth's rotation, when comparing both RCS and HCS, the RCS will constantly change its position (i.e., rotate) to keep pace with the radiant in the sky. It is precisely why the RCS is advantageous for modeling the meteor's path as, for this system, the meteor will not change the initial values of its $x$ and $y$ components over the entire burning duration (since its lifetime is usually less than one second). It means that for the RCS, the only thing that changes over time is the value of the $z$ component. Thus, the use of the RCS enables us to reduce abruptly the number of equations involved in the solution of this problem.

Moreover, notice that the conversion between the coordinates from the Horizontal System to the Radiant System is very straightforward. As this transformation only moves points without changing the distance between them, this conversion can be written by a rotation matrix $\mathbf{R_{zy}}$ in the form,

$$\mathbf{R_{zy}} = \mathbf{R_z}(\phi)\mathbf{R_y}(90° - \lambda), \tag{3.1}$$

where, as proposed by Goldstein et al. [6],

$$\mathbf{R_z}(\gamma) = \begin{bmatrix} \cos(\gamma) & \sin(\gamma) & 0 \\ -\sin(\gamma) & \cos(\gamma) & 0 \\ 0 & 0 & 1 \end{bmatrix} \tag{3.2}$$

and

$$\mathbf{R_y}(\beta) = \begin{bmatrix} \cos(\beta) & 0 & -\sin(\beta) \\ 0 & 1 & 0 \\ \sin(\beta) & 0 & \cos(\beta) \end{bmatrix} \tag{3.3}$$

which represents a counterclockwise rotation around the $z$-axis and the $y$ axis respectively.

To summarize, the Radiant Coordinate System is constructed using the Horizontal Coordinate System (which is a well-known celestial coordinate system) as reference. This is important because it allows us to identify which movements are necessary to convert the coordinates of a point in the Radiant System to the Horizontal System and vise versa. Hence, given the coordinates of any radiant in the Horizontal System, such that $\phi$ is the azimuthal angle and $\lambda$ is the altitude angle, initially the $x$-axis is rotated by $\phi$ degrees with respect to the $z$-axis. After that, the $z$-axis is rotated by $90 - \lambda$ degrees with respect to the $y$-axis, obtaining the Radiant System (see Fig. 3).

## 4 MODELING THE METEOR'S PATH

In this section, we aim to describe the meteor path. In this manner, we need to define which are the points where the meteor will appear and disappear in the sky. Considering a radiant of azimuth $\phi$ and altitude $\lambda$ visible to an observer on Earth (i.e., the radiant is above the horizon line) in a clear night sky. Given a meteor $\vec{m}$ which belongs to this radiant; if $\vec{m}$ is visible, then it is likely to be in the range of 80 to 120 $km$ from the Earth's surface, thus, its initial height is $h_i = [80, 120]$.

### 4.1 Starting time

Let $\vec{u}$ be the position in which the meteor arises, i.e., the initial position in which the meteor is visible to an observer. Initially, let us consider that the meteor is along the $z$-axis of the Radiant Coordinate System. In this way, from 2.1 we obtain the $z$ component of $\vec{u}$ for a given initial height $h_i$ with respect to the Earth's surface, that is,

$$u_z = \sqrt{r^2 \cos^2(\tfrac{\pi}{2} - \lambda) + 2rh_i + h_i^2} - r\cos(\tfrac{\pi}{2} - \lambda) \tag{4.1}$$

where $r$ is the Earth's radius and $\lambda$ is the altitude angle of the radiant. As discussed in Subsection 2.2, if the angle between the meteor and the zenith is less than 65°, it is more straightforward and convenient to use 2.2, which provides a good approximation if $90° - \lambda < 65°$. Thus,

$$u_z = \frac{h_i}{\cos(\frac{\pi}{2} - \lambda)}, \quad \text{if } \lambda > 25°. \tag{4.2}$$

Although the extension of the meteor tail always leads the observer to conclude that the meteor originated from a single point (the radiant), it is noteworthy that it does not mean that all meteors will always become visible from the same point. In fact, in a meteor shower, meteors usually arise at random positions around the radiant. Thus, our next step is to define values for the $x$ and $y$ components of the vector $\vec{u}$. For this purpose, considering that $(x,y)$ is a point radially displaced from the radiant, we use polar coordinates in the radiant plane, where we randomly select a value for the radius $p$ and another value to the angle $\theta$.

It is easy to notice that $\theta$ can assume any value from 0 to $2\pi$; however, the definition of a range for $p$ may not be trivial. As shown in 4, when a meteor moves around the $z'$-axis, it also changes the initial height $h_i$ which was used to calculate the $u_z$. Actually, this is an important mechanism to guarantee that the greater $p$, the lower the chances of this meteor lying in the range of 80 to 120 $km$ in height. Therefore, as it occurs in a real situation, the chances of an observer seeing a meteor far from the radiant ($z'$-axis) are much lower than when it is close to the radiant.

Moreover, as previously discussed, we know that the farther the radiant ($z'$ axis) is from the horizon, the lower the chances of the observer seeing the meteor with the naked eye. In this way, we assume that $p$ must not be greater than $u_z$. Thus, as the radiant approaches the horizon, the greater the range for $u_z$ and, consequently, the lower the chance of seeing the meteor. Therefore, we obtain $p = [0, u_z]$, which introduces another mechanism to regulate the effective ZHR.

It is noteworthy that this random selection of two uniformly distributed variables for $p$ and $\theta$ does not generate a uniform distribution in the disk. Actually, in this case, a greater number of points are concentrated in the center, which is exactly what we would expect from our model. Hence, we have that,

$$u_x = p \cos \theta \tag{4.3}$$
$$u_y = p \sin \theta \tag{4.4}$$

where $p = [0, u_z]$ and $\theta = [0, 2\pi)$. Finally,

$$\vec{u} = \begin{bmatrix} p \cos \theta \\ p \sin \theta \\ u_z \end{bmatrix}. \tag{4.5}$$

Note that as $max(u_z) = 120\ km$, in our model, the maximum distance between the observer and the meteor is about 170 $km$. This is a reasonable constraint for the model, since any meteor more than 170 $km$ away would not be visible with the naked eye. It is important to emphasize that in this model, the extremely bright meteors (fireballs) are not taken into account. In fact, fireballs can be seen at much higher distances, but they are rarely witnessed.

Regarding a computational implementation, after finding the vector $\vec{u}$, it may be useful not to compute the path of meteors which are above 120 $km$ in height or which arise below the horizon, because in both cases the chances of seeing them are negligible. Using the rotation matrix $\mathbf{R_{zy}}$ (3.1), it is possible to obtain $\vec{u}$ in the Horizontal Coordinate System ($\vec{u}'$), where,

$$\vec{u}' = \mathbf{R_{zy}}\vec{u}. \tag{4.6}$$

Considering that $\alpha$ is the angle formed by the meteor and the horizon, then

$$\alpha = \arcsin\left(\frac{u'_z}{\|\vec{u}'\|}\right). \tag{4.7}$$

Thus, it may be convenient to compute only the meteors in which $u'_z < 120$ $km$ and $\alpha > 0°$.

## 4.2 Ending time

Let $\vec{v}$ be the position in which the meteor disappear in the sky. As we already know the meteor's coordinates in the *xy* plane in the radiant system, which is the same throughout the meteor's path, $v_x$ and $v_y$ will assume the same values of $u_x$ and $u_y$ respectively. In this way, to obtain $\vec{v}$ we just need to find the value of $v_z$.

Analogous to the procedures used to obtain $u_z$, in this case, we can also use 2.1, but with $h$ equals to 80 $km$, which is, in general, the lowest height that a meteor occurs. Thus,

$$v_z = \sqrt{r^2\cos^2(\tfrac{\pi}{2} - \alpha) + 160r + 80^2} - r\cos(\tfrac{\pi}{2} - \alpha) \tag{4.8}$$

where $r$ is the Earth's radius and $\alpha$ is the angle formed by the meteor and the horizon (see Fig. 4). Also, it may be interesting to use the first order approximation form when the angle between the meteor and the zenith is less than 65°, then,

$$v_z = \frac{80}{\cos(\tfrac{\pi}{2} - \alpha)}, \quad if\ \alpha > 25°. \tag{4.9}$$

There are some meteors which are extremely bright and can be in activity for more time than usual. These meteors are rare and, in general, are noticed when the radiant seems to be very close to the horizon (Fig. 5). They can trace almost horizontal paths, which sometimes unable the observer to see the end of their burning. Thus, in order to include these objects in our model (for the sake of a computational implementation), we assume that when the radiant is between 0° and 1.5°, this kind of meteor may occur. In this way, to assure the observer will see this meteor crossing the sky, we just need to make $v_z$ be equal to $-u_z$.

Note that in both cases, it is important to make sure this object is really a meteor (and not a meteorite), that is, it never hits the observer nor the Earth's surface. Hence, we need to calculate the minimum distance between the observer and the meteor ($b$) for each one of these cases and check if the final height is greater than 80 *km*. In the general case, we see from Fig. 4 that,

$$b = \sqrt{p^2 + v_z^2}. \tag{4.10}$$

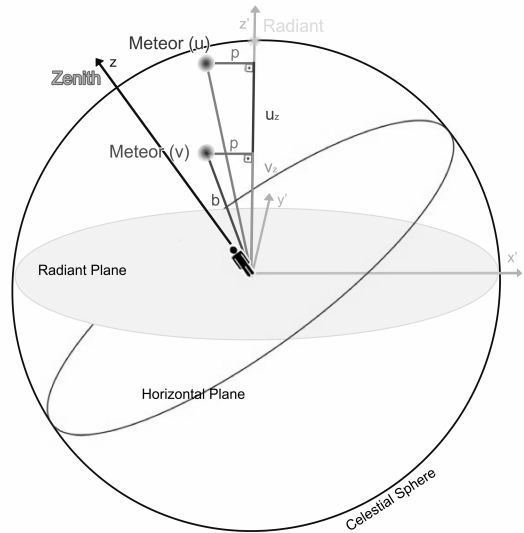

Figure 4: Modeling the meteor path. Finding the coordinates of the starting ($\vec{u}$) and ending ($\vec{v}$) time of the meteor trajectory in the Radiant Coordinate System, where $p$ is the projection of the meteor coordinates in the plane $x'y'$; and $b$ is the minimum distance between the observer and the meteor.

However, in the particular case where the radiant is close to the horizon, as illustrated in Fig. 5, we have $b = p$.

## 5 MODELING THE ZENITHAL HOURLY RATE

The activity period of a radiant may vary a lot. For example, it is expected that in 2018 the radiant *Draconids* is active from 6th to 10th October (5 days), on the other hand, it is expected that *Southern Taurids* is active from 10th September to 20th November (71 days). It means that during these days it may be possible to see meteors arising from these radiants [7].

As the Zenithal Hourly Rate (ZHR) refers only to the peak day, the effective ZHR usually decreases as the time from the peak day increases. As previously discussed, our meteor's path model already introduces a few mechanisms to make the effective rate be lower than the peak rate. However, we need to include a new mechanism to adjust the ZHR according to the time.

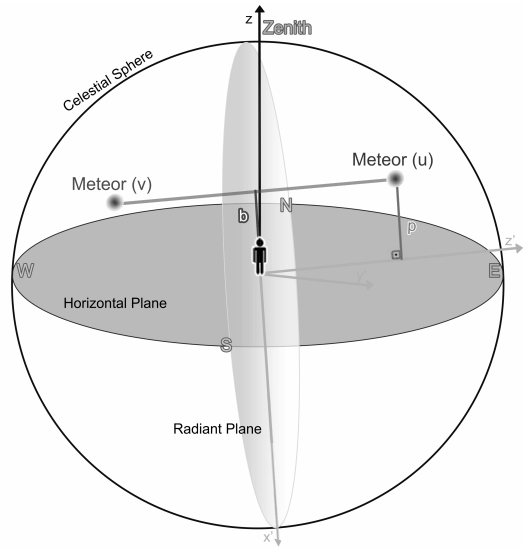

Figure 5: Modeling the meteor path. Finding the coordinates of the ending ($\vec{v}$) time for the particular case where the radiant is very close to the horizon.

Although we know that there are exceptions, such as the radiant having more than a peak day or the ZHR being almost constant during the whole activity period (when the ZHR is too low or when the activity period is too short), they rarely occur. Thus, for the computational implementation, we can assume that all meteor showers have only one peak day (at which the ZHR is maximal) and that the ZHR decreases exponentially as the time from the peak increases. In order to model this ZHR decrease, we use a Gaussian function in the form,

$$ZHR(t) = \kappa e^{-\frac{(t-k)^2}{2a^2}}, \quad i \leq t \leq f, \tag{5.1}$$

where $\kappa$ is the estimated ZHR in the peak day $k$, $t$ corresponds to the number of days in activity, $i$ and $f$ are, respectively, the first and last days of activity, and $a$ is defined as,

$$a = \begin{cases} \frac{k-i}{2} & if \quad i \leq t < k \\ \frac{f-k}{2} & otherwise \end{cases}. \tag{5.2}$$

For instance, Fig. 6 features the time-course of the ZHR for the radiant *Perseids*, which is active in 2018 from 17th August to 24th August with a ZHR of 110 meteors on 12th August [7]. It is important to mention that 5.1 may not produce a realistic estimation of the effective ZHR. Indeed, since it is a stochastic phenomenon, it would be impossible to determine this rate precisely using an analytical solution only with these basic data. Thus, Fig. 6 only illustrates what can be obtained from 5.1, which in turn provide a good estimation for our model.

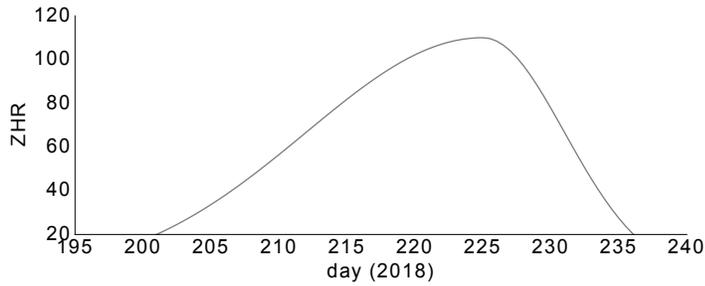

Figure 6: Zenithal Hourly Rate as a function of the time, in days, counted from 1st January 2018. The radiant *Perseids*, active from 17th July (199) to 24th August (237) with ZHR of 110 meteors in 12th August (225).

## 6   METEORS' SIMULATION

The model proposed in this paper was implemented as a new tool in *Stellarium*, which is a free and open-source planetarium software that aims to show a realistic three-dimensional sky for a given location and date. Moreover, it is also possible to simulate what an observer see with the naked eye or with some optical instrument.

Based on real catalogs, *Stellarium* is able to simulate and obtain real information about stars, planets, deep-sky objects, etc. In this context, our aim is to make this software able to simulate meteor showers and sporadic meteors. The model proposed in this paper was implemented and made available in the version *0.14.0* (and newer) of *Stellarium*, which can be downloaded from the official website of this software.

The data used to simulate meteor showers in *Stellarium* comes from the *International Meteor Organization* (IMO) and the *American Meteor Society* (AMS) catalogs, where it is possible to get information such as: activity, peak date, zenithal hourly rate (ZHR), radiant coordinates, radiant drift, meteor speed and population index.

As the meteor showers are (usually) annual events which occur in a specific period of time, in *Stellarium*, we define that each radiant can be in one of the three following states:

- **Inactive**, the radiant is inactive for the current sky date;
- **Confirmed**, the radiant is active, and its data was confirmed. Thus, this is a historical (really occurred in the past) or predicted meteor shower;
- **Generic**, the radiant is active, but its data was not confirmed. It means that this can occur in real life, but we do not have proper data about its activity for the current year.

The details of the software architecture as well as the physical analysis of the results obtained with our model are beyond the scope of this paper. However, we refer the interested reader to the supplementary material in [4] and the source code available at the official website (`http://stellarium.org`). The supplementary material [4] provides a few screenshots of the results that can be obtained in *Stellarium*. Namely, the figures illustrate how these three types of markers are displayed in the software; how the information about a specific radiant is printed on the screen and; how the user can search for active meteor showers in a given period of time. Also, we show a simulation of the night sky for an observer in the city of Rio de Janeiro, Brazil, in the peak day of the meteor storm of Leonids on 18th November 1833.

## 7   DISCUSSION

One of the most significant contributions of this work was the construction of a computational model capable of representing the meteor path for an observer located on Earth. In fact, the dynamics of a meteor to determine its height and the trail length are dependent on several variables, such as the entry speed, entry angle, atmospheric density, meteoroid mass and several other physical properties.

Although some work has considered such properties in the study of the dynamics of a meteor [2], to our knowledge, the simulation of meteor showers on the celestial sphere has not been studied to date. Moreover, in this paper, we propose a model that considers the case in which only basic information is known, such as the days of incidence, the zenithal hourly rate (ZHR) and the radiant position. Most papers about meteors aim to analyze and discuss observation techniques [10, 18, 21] and the relation between meteor showers and comets or asteroids [15, 9, 20].

This paper highlights the importance of choosing good reference systems and the power of a simple coordinate rotation. The Radiant Coordinate System (RCS) allows the conversion of a three-dimensional problem into a "unidimensional" problem, where only one of the three spatial coordinates needed to be updated over time. This approach led to a significant reduction in the complexity of the equations, which in turn enabled us to develop a plug-in for simulating meteors in *Stellarium*.

Regarding the implementation, the strategy of using an existing free and open-source software was a crucial aspect of this project. It simplified the software modeling process and enabled us to reach a large number of users quickly. A simple search on the Internet reveals that a wide range of websites has been using our software to show information about meteor showers. In addition, previous work [3] has discussed the use of the resources developed here (i.e., the meteor showers plug-in available in *Stellarium*) for the teaching of different physics concepts for high school students, including mechanics contents. Thus, we see that our work also has the potential to impact the teaching and learning of science.

To conclude, with this work, *Stellarium* became the first free software featuring such functions to date, enabling users to obtain real information on the incidence of meteor showers and its visibility for a given location and date. Future work will involve modeling fireballs; improving the meteor rendering; improving the data collection capabilities; and the exploration of realistic ways to model the meteor magnitude.

**RESUMO.** Neste trabalho, apresenta-se um modelo matemático que possibilita a simulação da trajetória de um meteoro como visto por um observador localizado em qualquer região na Terra. A estratégia de solução consiste em definir um novo sistema de coordenadas, denominado Sistema Radiante de Coordenadas, que é centrado no observador e tem seu eixo z alinhado com o radiante. Este novo sistema de coordenadas permite descrever a trajetória de um meteoro por meio de um número reduzido de equações, que por sua vez simplifica a solução do problema em questão. Posteriormente, este trabalho apresenta uma implementação computacional do modelo, que é incorporada como uma nova funcionalidade do Stellarium, um software livre e de código aberto. Ademais, discute-se que o modelo proposto pode ser utilizado tanto na simulação de chuvas de meteoros, quanto na simulação de meteoros esporádicos. Ressalta-se que as chuvas de meteoros são simuladas utilizando dados de catálogos reais.

**Palavras-chave:** modelagem computacional, meteoros, chuva de meteoros, Stellarium.